\documentstyle[11pt]{article}

\columnwidth\textwidth

\title{Microcanonical Quantum Statistics of Schwarzschild Black Holes}
\author{G\"unter Scharf \\ Institut f\"ur Theoretische Physik, \\ Universit\"at
Z\"urich, \\ Winterthurerstr. 190 , CH-8057 Z\"urich, Switzerland \\E-Mail:
Scharf@physik.unizh.ch} \date{} 
\begin{document} \maketitle

\def\d{\partial}\def\=d{\,{\buildrel\rm def\over =}\,}
\def\hq{\overline{\phantom{n}}\llap{h}}

\vskip 1.5cm {\bf Abstract:} It is shown that a quantized Schwarzschild black
hole, if described by a square root energy spectrum with exponential
multiplicity, can be treated as a microcanonical ensemble without problem
leading to the expected thermodynamical properties.

 \vskip 1.5cm Recently H.A.Kastrup [1-3] made
the interesting observation that a Schwarzschild black hole with energy spectrum
$E_n=\sigma\sqrt{n}E_P$, $(n=1,2\ldots, E_P=$Planck energy) and multiplicity
$g^n$ [4] leads to the expected thermodynamical properties if the divergent
canonical partition function $Z$ is analytically continued [2]. The analytic
continuation in the complex $g$-plane gives a complex $Z$, the desired
thermodynamics is obtained from its imaginary part. This situation has some
similarity with Langer's nucleation theory [5]. Therefore, it is argued [3] that
the complex canonical partition function might signal the instability of the
black hole. On the other hand, it is known that the statistical ensembles are
not always equivalent. In this case it is necessary to return to the
microcanonical ensemble which has the widest range of validity. For a black hole
this is even more appealing because it is hard to imagine a thermal bath coupled
to it. We will quickly see that our system can be treated microcanonically
without problem. This is even the simplest microcanonical calculation I know
because it uses mathematics from high school, only.

Let energy eigenvalues $E_n$ with multiplicities $\nu(n)$ be
$$E_n=b\sqrt{n},\quad\nu(n)=g^n,\quad n=1,2,\ldots\eqno(1)$$ where $b=\sigma
E_P$, $\sigma=$O(1), $g>1$ and $E_P=\sqrt{\hq c^5/G}$ is Planck's energy. The
microcanonical partition function $\Omega (E)$ is equal to the number of
eigenvalues below the energy $E$, hence
$$\Omega(E)=\sum_{n=1}^{n(E)}g^n={g^{n(E)}-1\over g-1},\eqno(2)$$ where
$$n(E)={E^2\over b^2}\eqno(3)$$ according to (1). To find the thermodynamics we
have to compute $$\omega={\d\Omega\over\d E}={2\log g\over g-1}{E\over
b^2}g^{E^2/b^2}.  \eqno(4)$$ Then the microcanonical temperature is given by
$$k_BT={\Omega\over\omega}={g^{E^2/b^2}-1\over g^{E^2/b^2}}{b^2\over 2E\log
g}.\eqno(5)$$ Its inverse is equal to $$\beta={2E\over b^2}\log g\Bigl[1+{\rm
O}\Bigl(g^{-E^2/b^2}\Bigl) \Bigl].\eqno(6)$$ Substituting for $E$ the rest
energy $E=Mc^2$ and [2] $$\sigma^2={\log g\over 4\pi},\eqno(7)$$ we get
Hawking's temperature $$\beta_H={8\pi Mc^2\over E_P^2},\eqno(8)$$ up to the
exponentially small corrections in (6). One can turn the thing around and leave
$n(E)$ in (2) unspecified. Then one finds $\beta= n'(E) \log g$. If one assumes
that $\beta\sim Mc^2=E$, one derives the square root spectrum $n(E)\sim E^2$,
$E\sim\sqrt{n}$.

The entropy is given by $${S\over k_B}=\log\Omega={\log g\over b^2}E^2-\log
(g-1)+{\rm O} \Bigl(g^{-E^2/b^2}\Bigl).\eqno(9)$$ Neglecting the constant second
term and the exponentially small correction we have $${S\over k_B}={\log
g\over\sigma^2}\biggl({E\over E_P}\biggl)^2= {\log
g\over\sigma^2}\biggl({Mc^2\over E_P}\biggl)^2.\eqno(10)$$ If we insert Planck's
energy $$E_P=\sqrt{\hq c^5\over G},$$ and the above value (7) of $\sigma^2$ we
find $${S\over k_B}={4GM^2\over\hq c}={A\over 4\pi l_P^2},\eqno(11)$$ where
$l_P^2=\hq G/c^3$ is the Planck length squared and $A=4\pi R_S^2= 16\pi
G^2M^2/c^4$ is the area of the horizon ($R_S=$ Schwarzschild radius).

The surprising consequence of these very simple results is that the black hole
looks like an innocent microcanonical equilibrium state.

\end{document}